\documentclass[12pt, preprint]{aastex}

\shorttitle{Tile or Stare?}
\shortauthors{Nemiroff}

\begin{document}

\title{Tile or Stare? Cadence and Sky Monitoring Observing Strategies that 
Maximize the Number of Discovered Transients}

\author{Robert J. Nemiroff}
\affil{Dept. of Physics, Michigan Technological University, Houghton,
MI 49931}

\begin{abstract} 
To maximize the number of transients discovered on the sky, should sky-
monitoring projects stare at one location or 
continually jump from location to location, tiling the sky?  If tiling is 
preferred, what cadence maximizes the discovery rate?  As sky monitoring is 
a growing part of astronomical observing, utilized to find such phenomena as 
supernovae, microlensing, and planet transits, well thought out answers to these 
questions are increasingly important.  Answers are sky, source, and 
telescope dependent and should include information about the 
source luminosity distribution near the observation limit, the duration of 
variability, the nature of the dominant noise, and the magnitude of down and 
slew times.  Usually, a critical slope of the effective cumulative 
transient apparent luminosity distribution (Log N - Log S) at the limiting 
magnitude will define when ``tile" or ``stare" is superior.  For shallower 
slopes, when ``tile" is superior, optimal cadences and pointing algorithms are 
discussed.  For transients discovered on a single exposure or time-contiguous 
series of exposures, when down and slew times are small and the character of the 
noise is unchanged, the most productive cadence for isotropic power-law 
luminosity distributions is the duration of the transient -- faster cadences 
waste time re-discovering known transients, while slower cadences neglect 
transients occurring in other fields.  A ``cadence creep" strategy might find an 
optimal discovery cadence experimentally when one is not uniquely predetermined 
theoretically.  Guest investigator programs might diversify previously dedicated 
sky monitoring telescopes by implementing bandpasses and cadences chosen to 
optimize the discovery of different types of transients. Example analyses are 
given for SuperMACHO, LSST, and GLAST.
\end{abstract}


\section{Introduction}

Humanity has monitored the sky by eye at least as long as history has been 
recorded.  Ancient records include, for example, bright supernovae and bright 
comets.  The idea of automated machine monitoring of the nighttime sky can  
be traced to several independent origins. Significant early efforts occurred in 
the X-Ray and gamma ray bands, including the Vela satellites that discovered 
gamma-ray bursts, first reported in 1972 (Klebesadel, Strong, and Olson 1972).  
Particularly noteworthy was the Burst and Transient Source Experiment (BATSE) 
deployed on board the Compton Gamma-Ray Observatory from 1991-2002 that kept a 
continuous monitor of the entire sky in the gamma-ray band, discovering over 
2000 gamma-ray bursts and the phenomena now known as Terrestrial Gamma Flashes 
(TGFs).  A continuously changing armada of satellites and instruments continues 
to monitor the entire sky in the gamma-ray band (see, for example, 
\citealt{Cli99}). 

The idea of continuous machine monitoring in {\it optical} of large portions of 
the nighttime sky can also be traced to several independent origins.  
\citet{Pac96} discussed possible scientific returns from monitoring the entire 
optical sky to detect different forms of variability. Coincidentally, the GROSCE 
\citep{Ack93} project started automated epochal monitoring of a large fraction 
of the sky in September 1996, continuing on as the LOTIS (Park et al. 1997, 
Williams et al. 1997) and Super-LOTIS projects, which keep an archive.  
\citet{Nem99} discussed practical limitations of continuously monitoring and 
recording the entire sky.  The value of such a record would be the ability to 
discover transience at a later time, a possible advantage given large amounts of 
storage space and limited amounts of real-time computing power.  In 
\citet{Nem99}, distinctions were delineated between projects that continuously 
record the entire sky and epochal recording which involve observations that 
return to any one sky location only after a given epoch. 

Other notable projects that have monitored pieces of the optical sky include 
MACHO \citep{Alc93}, OGLE \citep{Uda92}, EROS \citep{Aub93}, 
AGAPE \citep{Ans97}, TASS \citep{Ric98}, ASAS \citep{Poj97, Poj98}, Stardial 
\citep{McC97}, and ROTSE \citep{Mar97}. In the infrared, a night sky monitor 
sensitive to (almost exclusively) cirrus clouds has been deployed to the Apache 
Point Observatory \citep{Hog01}.  Although this instrument returns images in 
near real time, almost no stars are visible. 

In the past few years, the number of sky monitoring projects has blossomed.  
Reasons for the increase in sky-monitoring popularity likely include dramatic 
increases in digital storage, transfer, and analysis capabilities, while the 
price for CCD cameras has continued to drop.  Reasons for sky monitoring include 
discovering distant supernovae, eclipsing binary stars, planetary occultations, 
Earth-crossing asteroids, distant comets, meteors, and microlensing.  A project 
list is maintained on a web page by Paczynski 
(\url{http://www.astro.princeton.edu/faculty/bp.html}).  An abridged version is 
given below as Table 1, edited to include three other programs deemed relevant.
These projects are best known by, and listed by, their acronym in column 1.  If 
a Principal Investigator could be identified, this person is listed in column 2.  
Typically, these monitoring projects do not have a paper, preprint, or even 
abstract published about their capabilities.  Significant information can be 
found from each of the project's web pages, however, and so this is listed in 
column 3 as it appeared in November 2002.

\begin{table}
\caption{Current Sky Monitoring Projects}
\begin{tabular}{clllllllllll}
\tableline\tableline
Project & PI & Web Page \\
\tableline
CONCAM &    Nemiroff, R. J. & \url{http://concam.net} \\
KAIT &      Filippenko, A. &  \url{http://astron.berkeley.edu/~bait/kait.html} 
\\
LINEAR &    LINEAR team	&     \url{http://www.ll.mit.edu/LINEAR/} \\
LONEOS &    Bowell, T. &      
\url{http://www.lowell.edu/users/elgb/loneos/} \\
MEGA &      Crotts, A. &      \url{http://www.astro.columbia.edu/~arlin/MEGA/} 
\\
NEAT &      Helin, E. &       \url{http://neat.jpl.nasa.gov/} \\
RAPTOR &    Vestrand, W. T. & \url{http://www.raptor.lanl.gov/} \\
Spacewatch & McMillan, R. S. & \url{http://spacewatch.lpl.arizona.edu/} \\
STARE &     Brown, T. M. &  
\url{http://www.hao.ucar.edu/public/research/stare/stare.html} \\ 
SuperMACHO & Stubbs, C.	& \url{http://www.ctio.noao.edu/~macho/} \\
TAOS &      Alcock, C. &      \url{http://taos.asiaa.sinica.edu.tw/index.html} 
\\
YSTAR &     Byun, Y.-I.	&     \url{http://csaweb.yonsei.ac.kr/~byun/Ystar/} \\
\tableline
\end{tabular}
\end{table}

Even more ambitious sky monitoring projects are being planned for future years.  
For brevity, only three example projects toward the high end of the cost 
spectrum are mentioned: LSST, Pan-STARRS, and GLAST, as listed below in Table 2.  

\begin{table}
\caption{Example Future Sky Monitoring Projects}
\begin{tabular}{clll}
\tableline\tableline
Project & PI & Web Page \\
\tableline
LSST &   Tyson, A. &    \url{http://www.lsst.org/} \\
Pan-STARRS & Kaiser, N. & \url{http://www.ifa.hawaii.edu/pan-starrs/} \\
GLAST &  Michelson, P. F. & \url{http://www-glast.stanford.edu/} \\
\tableline
\end{tabular}
\end{table}

The need for efficient sky monitoring pointing algorithms is therefore becoming 
increasingly important.  Sometimes the same sky monitor will use different 
pointing algorithms, exercising both a ``tile" and ``stare" mode (e.g. ROTSE: 
\citealt{Keh02}). Rarely, however, does a monitoring survey give detailed 
analysis 
explaining their chosen cadence or time allocation algorithm.  (The term 
``cadence" here will be taken to mean the average frequency of return to image 
the same field.)  This paper is therefore an attempt to begin a discussion of an 
attribute common to many sky-monitoring surveys -- a desire to maximize the 
number of transients discovered.  In Section 2 some background will be given 
discussing common to many sky-monitoring telescopes.  Section 3 will discuss 
maximizing quiescents while Section 4 will discuss some general principles of 
monitoring for transients including the case of transients where the luminosity 
distribution is described by a power law near and well below the survey limiting 
magnitude.  Section 6 will give some concluding discussion.

\section{Background}

A telescope will detect a source only if its signal peaks above the noise. 
For a source of apparent luminosity $l$ (here ``apparent luminosity" is used 
synonymously with ``flux"), the signal-to-noise 
ratio $S/N$ during a single observation can be described by 
\begin{equation} \label{SN}
S/N \propto { l t_e \over
\sqrt{l t_e + b t_e + c + d l^2 t_e} }
\end{equation}
where $t_e$ is the duration of the exposure, $b$ is the equivalent apparent 
luminosity  of the background skyglow that accumulates with exposure time (for a 
good discussion of detection rates in the face of backgrounds, see, for example, 
\citealt{Nem99}), $c$ is a constant background term not affected by 
exposure time, for instance read-noise, and $d$ is a site, time, telescope, and 
sky position dependent amplitude for scintillation noise (see, for example, 
\citealt{You67} and \citealt{Dra98}).

Given a fixed $S/N$ threshold, $b$, $c$, and $d$ at the source detection limit, 
equation (\ref{SN}) can be inverted to solve for $l_{dim}$ as a function of 
$t_e$. 
The type of noise that dominates observations divides important regimes in this 
function. Although the relation is expressible analytically, it will be 
convenient to write it as a single power law such that
\begin{equation} \label{tebeta}
l_{dim} \propto t_e^{\beta} .
\end{equation}
In a common case where $c$ and $d$ are small, $\beta$ can be found 
instantaneously at any $l_{dim}$ such that
\begin{equation}  \label{beta}
\beta = - {  l_{dim} + b \over
             l_{dim} + 2b } .
\end{equation}
Now if the background skyglow level $b$ is small compared to $l_{dim}$, 
$\beta$ tends toward $-1$ so that $l_{dim} \propto t_e^{-1}$.  The other extreme 
case is when background skyglow $b$ is high compared to $l_{dim}$, so that the 
$b$ term dominates, $\beta$ tends toward $-1/2$, and so $l_{dim} \propto t_e^{-
1/2}$.

A frequent assumption used in these analyses will be that that the effective 
cumulative apparent luminosity distribution of sources (``Log N - Log S", 
hereafter just referred to as ``brightness distribution") is a power law such 
that the number of interesting objects accumulated during a single exposure of 
duration $t_e$ would be simply $N \propto l_{dim}^{\alpha}$.  Non-power law 
brightness distributions can frequently be approximated by a power law at (and 
below) the apparent luminosity cut-off $l_{dim}$, an approximation that makes 
the following discussion particularly relevant to realistic programs.

At the limit of observation, sources may be so numerous that their point spread 
functions begin to significantly overlap. When this happens, source confusion 
will create a practical limit on the faintest source detectable.  In practice, 
source confusion can be incorporated into the above formalism by allowing it to 
change the brightness distribution $N(l)$ for the given object type, telescope, 
and sky survey.  In fact, since $N(l)$ is an {\it effective} distribution, a 
host of practical limitations can be incorporated into it.

\section{Counting Quiescents}

Although this paper is primarily interested in transients, it is relevant and 
instructive to analyze the simpler case of quiescent sources first.  A canonical 
telescope and camera is assumed, with a given field of view of solid angle 
$\Omega_{field}$, and limiting apparent luminosity $l_{dim}$ observable over the 
telescopes bandpass.  Following \citet{Pee93} and \citet{Hog00}, the number of 
observable quiescent sources that would be visible in a single field to limiting 
apparent luminosity $l_{dim}$, found during an exposure of effective duration 
$t_e$ would be
\begin{equation} \label{Nqint}
N_{quiescent} (l>l_{dim}) = \int_{z=0}^{\infty}
              \int_{L=L_{min}(l_{dim},z)}^{\infty}
              { \Phi(L,z) K(L,z) D_H (1+z)^2 D_A^2 
              \over
              \sqrt{\Omega_M (1+z)^3 + \Omega_k (1+z)^2 
              + \Omega_{\Lambda} } }
              \ \Omega_{field} \ dL \ dz ,
\end{equation}
where $z$ is redshift, $L$ is absolute luminosity (also sometimes known as 
intrinsic luminosity or just ``luminosity", although it may be normalized), 
$\Phi$ is the luminosity function of candidate quiescent sources, $K$ is the k-
correction for the telescope bandpass, $D_H=c/H_o$, $c$ is the speed of light in 
vacuum, $H_o$ is the Hubble parameter, $D_A$ is angular diameter distance, 
$\Omega_M$ is the energy density of the universe in matter in units of the 
critical energy density, $\Omega_k$ is the energy density due to the curvature 
of space, and $\Omega_{\Lambda}$ is the energy density due to the cosmological 
constant.   The integrand $L_{min}$ can be computed from $l_{dim}$ and 
$z$ from $L_{min}=l_{dim} D_L^2 / K(z)$ where $D_L$ represents 
luminosity distance. 

A perhaps familiar case is that of detecting quiescents in fields of such low 
background that Poisson noise dominates the counting statistics.  This will be 
referred to as the ``low background" case.  Given a canonical telescope and set 
amount of time $t_c$ for an observing campaign, should this time be divided 
tiling all available fields, or spent staring at a limited number of fields?  It 
will be assumed here that data taken from the same parts of the sky can be 
efficiently co-added.  

To make matters simple, it will be assumed that the brightness distribution can 
be simplified from equation (\ref{Nqint}) to $N_{quiescent} \propto 
l_{dim}^{\alpha}$.  From equation (\ref{tebeta}) $l_{dim} \propto t_e^{\beta}$ 
so that equation (\ref{Nqint}) simplifies to
\begin{equation} \label{NqPoisson}
N_{quiescent} \propto t_e^{\alpha \beta}. 
\end{equation}
Studying this simple equation will give significant insight into the ``tile" or 
``stare" question.  To calibrate intuition, let's consider the case of 
$\alpha \beta = 1$, so that the number of detected sources is just 
linear with the exposure time.  An example case is when the background is low, 
$\beta \sim -1$, and so $\alpha \sim -1$.  Assuming it takes little time to slew 
to a new field, it then does not matter if one stares at the same location, or 
tiles the sky: the same number of long duration (and hence quiescent) transients 
will be detected.  Here the answer to ``tile or stare" is a formal tie. 

If the power-law index $\alpha \beta$ is less than 
unity, however, equation (\ref{NqPoisson}) indicates that ``tile" will detect 
more sources per unit exposure time than ``stare".  This is because, when 
staring at a single field, new sources are appearing over the limiting 
brightness horizon at an increasingly slower rate.  Higher rates are found by 
starting over on a new field.  Given a total observing campaign time, the most 
sources will be found by dividing the time equally between all the available 
fields.

Similarly, if the power-law index $\alpha \beta$ of equation (\ref{NqPoisson}) 
is greater than unity, "stare" will detect more sources per unit exposure time 
than ``tile".  This is because, when staring at a single field, new sources are 
pouring over the limiting brightness horizon at an increasingly fast rate.  
Lower rates would be found by starting over on a new field.  Therefore, in 
general, given a total observing campaign time, the most sources will be found 
by staring at one field.

Stated differently, two identical observations of statistically identical 
fields should yield twice the number of sources than in a single field.  For a 
steep brightness distribution where $\alpha \beta > 1$, however, spending twice 
the time on the first field will detect {\it more} than twice the number of 
sources.  So ``stare" is preferred to maximize sources observed or monitored.

Can $\alpha > 0$?  Since the brightness distribution $N_{quiescent}$ is a 
cumulative distribution, it cannot decrease, an equivalent statement to not 
having $\alpha > 0$ over any part of its length.

For non-power law brightness distributions, the situation is a more complex.  
Cases likely include where the luminosity function is not a power law ($\alpha$ 
changes) and cases where noise terms are not constant over the course of 
observations ($\beta$ changes). For $\alpha \beta$ decreasing monotonically with 
increasing $t_e$ in identical fields, one should always observe in the field 
where the rate of source accumulation is instantaneously highest.  Therefore, 
one should stare at one field only until the rate of source accumulation falls 
below that in a fresh field.  This is certainly true when $\alpha \beta$ falls 
through unity, although the transition will likely occur in many cases when 
$\alpha \beta$ is still in excess of unity.

Note that the tile/stare divide for brightness distributions well-characterized 
by a power-law $\alpha$ is $\alpha = 1/\beta$ which depends on the level of 
the background at the limiting apparent luminosity of the single field exposure 
$l_{dim}$.  When sky flux $b$ is negligible, a case here referred to as ``low 
background", equation (\ref{beta}) indicates that $\beta = -1$ so that the 
divide in terms of the brightness distribution comes at $\alpha = 1/\beta = -1$.  
When the sky flux dominates, a case here referred to as ``high background", then 
$\beta = -1/2$ so that the divide in terms the brightness distribution comes at 
$\alpha = -2$.  In general, the critical brightness distribution power-law index 
at the tile/stare divide is
\begin{equation} \label{alphacritical}
\alpha_{critical} = 1 / \beta = -{ l_{dim} + 2 b \over
                                 l_{dim} + b } .
\end{equation}
When $\alpha > \alpha_{critical}$, tiling will optimize source counts, otherwise 
staring will optimize source counts.

The longer an instrument observes a particular field, the fainter the source 
detectable at the limiting apparent luminosity ($l_{dim}$ decreases), the 
smaller the limiting source brightness will be compared to the background sky 
brightness. Stated differently, even if a field observation started at ``low 
background", it naturally migrates toward ``high background" as exposures 
lengthen.  This means that for long exposures, $\beta$ naturally migrates toward 
higher values, so that $\alpha_{critical}$ will migrate toward a lower value.  
Therefore, a switch from ``stare" to ``tile" might become advantageous even were 
the brightness distribution power-law $\alpha$ to remain constant.

For non-power law, non-monotonically decreasing brightness distributions, 
maximizing source counts becomes similar to a chess game.  Fields that start 
with low source accumulation rates might ramp up quickly at a later time, when, 
for example, a cluster might become resolved.  Therefore, choosing which field 
to image next and for how long in order to maximize source counts might require 
a complex Monte-Carlo program, possibly one that operates in real time including 
topical information about how seeing and weather affect the (effective apparent) 
brightness distributions in fields across the sky.

\section{Counting Transients} 

In this paper a practical distinction will be made between quiescent sources and 
long duration transients.  Here, increasingly faint quiescent sources can be 
detected by co-adding images of the same part of the sky at any time, whereas 
transients will need to be detected on a single exposure or co-added 
time-consecutive series of sky exposures.  For transients of any duration, the 
fleeting nature of the source makes the ``tile" or ``stare" question more 
complicated when trying to maximize the discovery rate.

If telescope fields are easily aligned and relevant data are easily available, 
it will be possible to discover transients on time-separated exposures, possibly 
by co-adding frames taken at different times during the transient.  For 
simplicity, however, only the relatively standard paradigm of discovering 
transience in a time-contiguous series of exposures will be considered here. 

Discovering a transient with a single exposure is particularly susceptible to 
false triggers by non-astronomical phenomena.  Transience verification is 
usually necessary for a practical sky-monitoring algorithm.  False triggers are 
usually a single-frame phenomenon, however, and reality verification can be 
built into a time-contiguous series of exposures. When these check observations 
occur time-contiguous with the initial observation, together they can be 
considered as part what is necessary for transient ``discovery."

Given that ``tile" is desired, the tiling cadence should of course be chosen for 
what science return is expected.  In general, a ``discovery cadence" will be 
distinguished from a ``tracking cadence."  Discovery cadence, for example, 
should maximize the number of transients discovered.  Tracking cadence, however, 
should maximize the scientific return from a single transient.  It is 
possible -- even likely -- that a non-uniform cadence would better address both 
discovery and tracking for those monitors not reporting triggers to follow-up 
instruments.  However, unless explicitly stated, uniform cadence rates will be 
assumed in this paper.  In addition, in this paper an attempt will be made to 
maximize the number of discovered transient objects, expecting that important 
transient event will be handed off (in a timely fashion) to a telescope 
dedicated to following them up specifically.

The way the apparent brightness distribution is defined for quiescents and 
transients might differ.  In particular, $l_{dim}$ for a transient used in 
equation (\ref{Ntint}) can be defined a number of ways.  Useful definitions 
include $l_{dim}$ as the apparent luminosity during a quiescent phase, as the 
apparent luminosity at the peak of an outburst, or as the average apparent 
luminosity over a given duration.  Given a corresponding distance, apparent 
luminosity $l$ and absolute luminosity $L$ can be directly related.

Each transient will have amplitude $A$, which can run from less than unity 
(sources become dimmer, such as during a planetary transit) to greater than 
unity.  For explosive sources, of course, the $A$ is expected to be much greater 
than unity.  Transients will have an amplitude probability density such that the 
probability of a transient of absolute luminosity $L$ and redshift $z$ having an 
amplitude between $A$ and $A + dA$ at any time is given by $\psi(A,L,z) dA$.  
For each $L$ and $z$ this probability is normalized to unity so that 
$\int_0^{\infty} \psi(A,L,z) dA = 1$.

Similarly, each transient will have a duration of $t_{dur}$.  Transients 
will have a duration probability density such that the probability of a 
transient of absolute luminosity $L$ and redshift $z$ having an duration between 
$t$ and $t + dt$ at any time is given by $\xi(t,L,z) dt$.  This probability is 
also normalized to unity so that $\int_0^{\infty} \xi(t,L,z) dt = 1$.  It is 
assumed here that all $t_{dur} >> t_e$, so that these terms do not appear 
explicitly in the following analyses. 

Including these factors, the number of transients detected in a single field 
down to limiting apparent luminosity $l_{dim}$ in a single exposure (or a single 
consecutive series of exposures) of (total) duration $t_e$ would be
\begin{eqnarray} \label{N1tint}
 \nonumber       N_{transient} (l>l_{dim}) = 
                 \int_{z=0}^{\infty}
                 \int_{A=0}^{\infty}
                 \int_{L=L_{min}(l_{dim},A,z)}^{\infty} \\
                 {\Phi(L,z)
                 \psi(A,L,z)
                 K(L,z) D_H
                 (1+z)^2 D_A^2 
                 \over
                 \sqrt{\Omega_M (1+z)^3 + \Omega_k (1+z)^2 
                 + \Omega_{\Lambda} } } \ \Omega_{field} \ dL \ dA \ dz .
\end{eqnarray}
Here $N_{transient}$ is {\it not} a rate -- it is the number of transients 
discovered during exposure time $t_e$, assuming each transient's source counts 
accumulate during the entire $t_e$.  During exposure, the sky is considered to 
be static in the sense that extending $t_e$ will not increase the number of 
bright transients, but allow more faint transients (visible during the entire 
exposure) to come over the detection threshold. In this light, $\Phi(L,z)$ is 
also {\it not} a rate and describes the space density of transients that happen 
to be observable at absolute luminosity $L$ and redshift $z$ when integrated 
over the entire length of the exposure.  Although $t_e$ does not appear 
explicitly in equation (\ref{N1tint}), it will enter later through $l_{dim}$.

Equation (\ref{N1tint}) has two realizations involving $L_{min}$, the minimum 
absolute luminosity of a transient that is seen at the apparent detection 
threshold $l_{dim}$.  When known quiescent sources are being inspected for 
transience, $L_{min}$ is determined as it was below equation (\ref{Nqint}).  
It is possible, however, that transient sources will be detected {\it only} 
because they show transience.  This might happen, for example, were a 
quiescent source originally below the detection threshold of the telescope to 
experience a high amplitude event that brought it above the detection threshold.  
This would affect the minimum absolute luminosity in equation (\ref{Ntint}) such 
that $L_{min} K(z) A = l_{dim} D_l^2$.  An example of this type of transient 
detection is pixel lensing \citep{Cro92,Gou95}.

When re-inspecting a given field for transience, the probability of finding new 
transience may have changed.  Suppose, for example, that the same field is 
inspected twice for transience, one time shortly after the other.  The two 
exposures are not co-added.  Given that the exposure time $t_e$ is the same for 
each field but significantly less than transient duration $t_{dur}$, the 
likelihood of finding {\it new} transients in the second exposure is likely 
reduced.  This is because there has only been a short time during which a new 
transient could have gone off.

More generally, if $N_1$ is the number of transients discovered in 
the first field exposure, and the same field is inspected for an equal exposure 
time $t_e$ after time $t_{return}$, then $N_2$, the number of transients 
discovered in the second exposure is related to $N_1$ by
\begin{equation}
                 N_2 = {\rm min} (t_{return}/t_{dur},1) N_1 .
\end{equation}
It is assumed that the transient rate remains constant in each region of space.  
Note that as $t_{return}$ becomes small, so does the number of new transients 
discovered.  When $t_{return}$ becomes larger than $t_{dur}$, then $N_2 = N_1$ 
so that field has effectively been ``reset" and all discovered episodes of 
transience at duration $t_{dur}$ are again new.

The number of new transients discovered in a previously observed field is 
therefore given by
\begin{eqnarray} \label{Ntint}
 \nonumber       N_{transient} (l>l_{dim}) = 
                 \int_{z=0}^{\infty}
                 \int_{A=0}^{\infty}
                 \int_{L=L_{min}(l_{dim},A,z)}^{\infty} \\
 \nonumber       {\Phi(L,z)
                 \psi(A,L,z)
                 K(L,z) D_H
                 (1+z)^2 D_A^2 
                 \over
                 \sqrt{\Omega_M (1+z)^3 + \Omega_k (1+z)^2 
                 + \Omega_{\Lambda} } } \\
                 \Omega_{field} \
                 {\rm min} (t_{return}/t_{dur}(L,z,A),1)
                 \ dL \ dA \ dz .
\end{eqnarray}

Equation (\ref{Ntint}) has realizations that are quite complicated.  As with the 
general quiescent case, for non-power law, non-monotonically decreasing 
transient brightness distributions, maximizing source counts becomes similar to 
a chess game.  Choosing which field to image next and for how long in order to 
maximize discovered transients might require a complex Monte-Carlo program, 
possibly one that operates in real time including topical information about how 
seeing and weather affect the apparent brightness distributions in fields across 
the sky.  Significant insight can be gleaned, however, from studying relatively 
simpler theoretical regimes.

\subsection{Isotropic Power-Law Brightness Distributions}

In the cases considered here, fields are identical, noise sources are unique and 
unchanging over observations of interest ($\beta$ is a constant), and 
the effective apparent brightness distribution is a single power law over its 
entire length ($\alpha$ is a constant).  Additionally, it will be assumed that 
transients attain amplitude greater than $A$ for duration $t_{dur}$.  
Mathematically, this means $\psi(A') = \delta(A'-A)$ in equation (\ref{Ntint}). 

Suppose an observing campaign is limited to a set amount of total observing time 
$t_c$. Given that the time is divided between $N_{field}$ identical fields, the 
time between returning to re-image the same field is given by
\begin{equation} t_{return} = N_{field} (t_e + t_d)
                                  + (N_{field} -1) t_s
              \approx N_{field} (t_e + t_d + t_s)
\end{equation}
where $t_d$ is the down time it takes for the imager to reset before taking 
the next image, and $t_s$ is the time it takes to slew to the next field.  The 
approximation is true when $N_{field} >> 1$. Note that $t_d$ will likely include 
the time it takes to read out the data and take dark frames. 

Given that $t_c$ is large compared to a single exposure time, the total number 
of exposures taken in the campaign will be 
\begin{equation}  \label{Nc}
 N_c = {t_c \over t_e + t_d + t_s } .
\end{equation}
The grand total number of sources observed in the campaign will then be
\begin{equation}  \label{NcNt}
 N_g = N_c N_{transient}.
\end{equation}
Note that these equations are valid even for dedicated telescopes -- $t_c$ is 
then proportional to the lifetime of the telescope.

The number of transients discovered upon return to a single 
field can then found from equation (\ref{Ntint}), which simplifies to
\begin{equation} \label{Nt}
N_{transient} \propto t_e^{\alpha \beta} 
                     \ {\rm min} (t_{return}/t_{dur},1) .
\end{equation}

Therefore the grand total number of transients discovered during the campaign is 
found by combining Eqs. (\ref{Nt}) with equation (\ref{Nc}) and equation 
(\ref{NcNt}) so 
that 
\begin{eqnarray} \label{Ng_beta}
N_g \propto { t_e^{\alpha \beta} \over t_e + t_d + t_s } \
            ( t_{return} > t_{dur} ) \\
N_g \propto { t_e^{\alpha \beta} } \
            ( t_{return} < t_{dur} ) . \\ \nonumber
\end{eqnarray}
Stated explicitly for the extreme cases of high and low background,
\begin{eqnarray}   \label{Ng}
N_g \propto  { t_e^{-\alpha} \over t_e + t_d + t_s } \
             (t_{return} > t_{dur}, \ {\rm low\ background}) \\
N_g \propto  { t_e^{-\alpha/2} \over t_e + t_d + t_s } \
             (t_{return} > t_{dur}, \ {\rm high\ background}) \\
N_g \propto   t_e^{-\alpha} \
             (t_{return} < t_{dur}, \ {\rm low\ background}) \\
N_g \propto   t_e^{-\alpha/2} \  
             (t_{return} < t_{dur}, \ {\rm high\ background}) . \\ \nonumber
\end{eqnarray}
For the cases where $t_{return} > t_{dur}$ one can set $dN_g/dt_e=0$ and 
solve for $t_e$ yielding
\begin{equation}
t_e^{best} = { \alpha \beta (t_d + t_s) \over 1 - \alpha \beta } .
\end{equation}

Note that these equations are consistent with many of the conclusions derived 
above for quiescent sources. When $\alpha \beta$ nears $1$, the denominator 
nears zero so that the $t_e^{best}$ goes to infinity, indicating that ``staring" 
is becoming better than tiling.  When $\alpha \beta$ approaches $0$, the 
numerator and hence $t_e^{best}$ also approach zero, indicating an increasingly 
rapid tiling rate.

Why can't a $t_e^{best}$ be found for cases when $t_{return} < t_{dur}$?  
Solving $dN_g/dt_e = 0$ formally indicates that $N_g$ is best maximized for the 
longest values of $t_e$. Therefore the minimum critical cadence rate is 
$t_{return}^{critical} = t_{dur}$ so that 
\begin{equation}
 t_e^{critical} = { t_{dur} \over N_{field} } - t_d - t_s .
\end{equation}

When $t_{return} \ge t_{dur}$ the actual ``best" cadence rate that maximizes the 
number of transients discovered can be found by substituting the above 
$t_e^{best}$ equations into $t_{return} = N_{field} (t_e + t_d + t_s)$ which 
become
\begin{equation}  \label{treturn}
t_{return}^{best} = {\rm max} ( t_{dur}, 
N_{field} { t_d + t_s \over 1 - \alpha \beta } ) .
\end{equation}

The situation is shown graphically in Figures 1-4. These figures show plots of 
$N_g$ versus $t_{return}$.  Here $N_g$ is normalized to the number of transients 
discovered when $t_{return} = t_{dur}$.  The value of $N_{field}$ is taken to be 
large compared to unity, while $t_c$ is taken to be large compared to all other 
durations.  Their precise values are not important and do not affect the plots.  
On each of the Figures three curves are drawn, corresponding to $\alpha$ values 
of $-2.5$, $-1.5$, and $-0.5$.

Figure 1 depicts the transient discovery rate during low background ($\beta = -
1$) and negligible down and slew times.  In other words, Poisson counts dominate 
the noise and $t_d + t_s << t_{dur}/N_{field}$.  Formally, $t_d + t_s = 0$.
Inspection of Figure 1 indicates that for a shallow brightness distribution such 
as $\alpha = -0.5$, ``tile" is the most productive strategy, and the cadence 
that maximizes transient discovery is $t_{dur}$, the duration of the transient.  
For steep brightness distributions such as $\alpha = -1.5$ and $\alpha = -2.5$, 
longer return rates result in a greater number of transients recovered, 
indicating that ``stare" is the best policy.  From the above analysis, note that 
$\alpha = -1$ is the dividing line between ``tile" and ``stare."

Figure 2 similarly depicts the transient discovery rate during high background 
($\beta = -1/2$) and negligible down and slew times.  In other words, a constant 
sky background dominates the noise while $t_d + t_s << t_{dur}/N_{field}$.  
Formally, again, $t_d + t_s = 0$. Inspection of Figure 2 indicates that for 
shallow brightness distributions such as $\alpha = -0.5$ and $\alpha = -1.5$, 
``tile" is the most productive policy, and the cadence that maximizes transient 
discovery is $t_{dur}$, the duration of the transient.  For the steepest 
brightness distribution $\alpha = -2.5$, slower cadences result in greater 
transients recovered, indicating that ``stare" is the best policy.  From the 
above analysis, note that $\alpha = -2$ is the dividing line between ``tile" and 
``stare."

Figure 3 similarly depicts the transient discovery rate during low background 
and significant down and/or slew times.  Specifically, the case where $t_d + t_s 
= (2/3) (t_{dur}/N_{field})$ is assumed.  Inspection of Figure 3 indicates that 
for a shallow brightness distribution such as $\alpha = -0.5$, ``tile" is again 
the most productive policy.  However, now the best cadence is slightly longer 
than $t_{dur}$, and is given by equation (\ref{treturn}). For steeper brightness 
distributions such as $\alpha = -1.5$ and $\alpha = -2.5$, longer return rates 
again result in greater transients recovered, indicating that ``stare" is the 
best policy.  From the above analysis, note that $\alpha = -1$ is the 
dividing line between ``tile" and ``stare."

Figure 4 similarly depicts the transient discovery rate during high background 
and significant down and/or slew times. Specifically, the case where $t_d + t_s 
= (2/3) t_{dur}/N_{field}$ is assumed.  Inspection of Figure 4 indicates that 
for shallow brightness distributions such as $\alpha = -0.5$ and $\alpha = -
1.5$, ``tile" is the most productive policy, and the cadence that maximizes 
transient discovery is given by equation (\ref{treturn}). For the steepest 
brightness 
distribution $\alpha = -2.5$, slower cadences result in greater transients 
recovered, indicating that ``stare" is again the best policy.  From the above 
analysis, note that $\alpha = -2$ is the dividing line between ``tile" and 
``stare."

Transient durations can be so short that they are less than the 
exposure time.  In that case, an effective apparent luminosity should be used 
that incorporates the amount of actual integrated light over the entire 
exposure, instead of the peak apparent luminosity of the transient.  A more 
complicated paradigm not considered here are transients with durations longer 
than $t_{return}$ that are detected by co-adding counts during each 
return exposure.

The results of this section can be summed up as follows: If, during exposure, 
the rate that transients come over the limiting magnitude horizon is increasing 
fast enough ($\alpha \beta > 1$), then ``stare" is preferred.  If, on the 
other hand, the rate that transients come over the limiting magnitude horizon is 
not increasing fast enough ($\alpha \beta \le 1$), then ``tile" should be 
preferred.  Usually the best tiling cadence is the duration of the transient, 
since a faster tiling cadence will waste effort on transients that have been 
previously discovered, while a slower tiling cadence will miss transients 
occurring in other fields.  If, however, the duration of the transient is 
comparable to the cumulative read-out and/or slew times during a sky-tiling, 
then a mathematical maximization as described above in equation (\ref{treturn}) 
will 
find the most productive cadence.

\section{Example Applications}

\subsection{Local Uniform Isotropic Standard Candle Quiescents}

Perhaps the most intuitive example is also the most instructive: that of uniform 
and isotropic standard candles in a local Newtonian universe.  Given that 
cumulative source numbers increase as the cube of their distance and their 
apparent luminosity falls as the square of their distance, many an introductory 
text book correctly states that $N_{source} \propto l_{dim}^{-1.5}$, meaning 
that $\alpha = -1.5$.  

Suppose further that these sources are quiescents and that a campaign of time 
$t_c$ on a given telescope is dedicated to observing as many of them as 
possible.  Is ``tile" or ``stare" the best observing strategy?  As indicated 
above in Section 4, the answer is ``tile" if $\alpha \beta < 1$, and ``stare" 
otherwise.  Since $\alpha = -1.5$, $\beta$ becomes the determining factor.  If 
$\beta < -2/3$, tiling will maximize quiescent counts, otherwise staring will. 
Note that this $\beta$ is between the above discussed cases of low and high 
background, so that when the background is low, $\beta = -1$, and staring is the 
best strategy.  Alternatively, when the background is high, $\beta = -1/2$, so 
that tiling wins.  Since quiescents can be discovered in frames co-added at any 
delay, the exact tiling rate is not important, and so can be set to minimize the 
total slew time, for example.  A tiling campaign should best proceed by dividing 
the time equally between observable identical fields, with fields having the 
least exposure time getting the highest priority.

\subsection{Local Uniform Isotropic Standard Candle Transients}

Suppose now that the above uniform, isotropic sources are transients with 
duration $t_{dur}$ and characteristic apparent luminosity $l$.
All transients will be assumed to have the same duration. The fraction of 
sources that show transience at any one time will turn out to be unimportant for 
optimization.  Again consider the search a campaign of time $t_c$ on a given 
telescope. 

Here, again, the ``tile" or ``stare" decision depends on the predominant source 
of noise.  Again ``stare" will be preferred when $\alpha \beta > 1$, equivalent 
to $\beta < - 2/3$ since $\alpha = -1.5$.  In the quiescent case, cadence was 
not important since sources could be discovered on fields co-added with any time 
delay.  Here the finite duration of transience will indicate a best cadence.
Suppose first that $t_{return} < t_{dur}$.  Equation (\ref{Ng}) indicates that 
$N_g$ increases monotonically with $t_e$, pushing us into the regime where 
$t_{return} \ge t_{dur}$.  Equation (\ref{Ng}), however, has $N_g$ {\it 
decreasing} monotonically with $t_e$ at large $t_e$.  Equation (\ref{treturn}) 
then gives $t_{return}^{best}$.  The best cadence is seen to be $t_{dur}$ for 
small down and slew times.

\subsection{Maximizing Microlensing With SuperMACHO}

The SuperMACHO project inspects the LMC for microlensing \citep{Stu02}.  
The LMC, however, shows an anisotropic and non-uniform sky distribution, 
indicating that the above detailed cadence calculations made for isotropic, 
uniform distributions are of mainly didactic value.  An analysis of the 
SuperMACHO observing algorithm is given by \citep{Gou99}.  According to their 
web page, SuperMACHO employs 60 fields each having $\Omega_{field} = 0.36$ 
deg$^2$.  The web page also indicates a canonical magnitude limit of around 
$V=23$.  It will be assumed that high background dominates the noise in any 
exposure, so that $\alpha_{critical} = -2$. 

According to Figure 4 of \citet{Alc00}, a canonical LMC field (Field 13 in their 
work) has a cumulative luminosity function where $N_g \sim l_{dim}^{-0.9}$ from 
visual magnitude 20 to visual magnitude 22, and $N_g \sim l_{dim}^{-0.6}$ from 
visual magnitude 22 to visual magnitude 24.  A combined average power-law index 
from 20 to 24 is about $\alpha=-0.7$.

Since the LMC stars are all at approximately the same distance, the 
cumulative luminosity function will be the effective cumulative apparent 
brightness distribution.  Therefore, since $\alpha > \alpha_{critical}$, the 
above analysis indicates that ``tile" will discover more transients than 
``stare."  Now a canonical duration of a microlensing event is about one month.  
To obtain good coverage, however, one might want to record the event on the 
rise, so a duration of interest is about two weeks.  Since $t_d$ and $t_s$ are  
on the order of seconds, it will be assumed that they are negligible compared to 
$t_{dur}/N_{field}$ and $t_e$.  The above analysis then indicates that for each 
field, the optimal $t_{return}$ time is $t_{dur}$.

This indicates that SuperMACHO should return to each field after two weeks.  The 
SuperMACHO web page notes, however, that each field is returned to twice a 
night, ``in order to maximize the number of stars inspected for microlensing."
Given that each star has a constant probability of being microlensed, the above 
quoted maximization scheme of maximizing stars would also maximize the 
transients discovered. Therefore, how can these two cadences be consistent?

One reason may be that the effective cumulative apparent luminosity distribution 
($N_{transient}$) of LMC transients is dropping rapidly after a given exposure 
time (C. Stubbs 2003, private communication).  The optimized cadence of two 
weeks assumed that a constant $\alpha \beta$ continued indefinitely. 

Now since each LMC SuperMACHO field is different, $N_g$ is likely different for 
each field.  As indicated above, for complex cases like these, a real-time 
Monte-Carlo routine might be run planning each night's observing campaign based 
on present and predicted sky conditions that could best maximize $N_g$ for that 
night.

\subsection{Maximizing Type IA Supernovae Discovered with LSST}

Suppose one wants to maximize the number of Type IA supernovae discovered with 
the planned Large-aperture Synoptic Survey Telescope \citep{Ang01}.  According 
to modeling in a Simple Cold Dark Matter universe by \citet{Por00}, the integral 
number count rate of these transients is approximately $N_g \propto l_{dim}^{-
2}$ for $21 < I < 24$, while $N_g \propto l_{dim}^{-0.5}$ for $24 < I < 27$.  
Now the LSST web page states a design goal of magnitude 24 in a single 10 second 
exposure over 7 deg$^2$, with a readout time is estimated to be about 5 seconds. 
Further suppose that LSST can tile 25\% of the sky per night ($\pi$ steradians; 
$\sim$ 10,000 deg$^2$).  This indicates that on a clear moonless night that 
$N_{field} \sim 1400$, LSST can point to about $1400$ independent fields.  A 
supernova might be perceived to have the most value if caught on the rising part 
of its light curve, which has duration of about $t_{dur} \sim 15 (1+z)$ days.  

This case is simpler than the SuperMACHO/microlensing case since Type IA 
supernovae can be assumed distributed isotropically in the universe.  
Also, supernovae should not crowd each other on the sky, so that we would not 
expect source confusion to flatten the effective brightness distribution.

At brighter magnitudes, the steep $\alpha = -2$ brightness distribution would 
place $\alpha \beta > 1$ for any $\beta$, indicating that LSST should stare at 
any field until the brightness distribution breaks.  At fainter magnitudes, the 
shallow $\alpha$ indicates that $\alpha \beta < 1$ for any $\beta$, indicating 
that LSST should tile in this regime.  

Given that $\alpha \beta$ is indeed a constant in this regime, the analysis 
given above in Section 4 can determine the most productive cadence.  Assuming 
the down and slew times are small, the cadence should be the duration of the 
interesting part of the transient: 15 days for a low-redshift supernova.  Given 
1400 fields and a 25 percent duty cycle due to the Sun and Moon, the best 
exposure time comes out to be about 230 seconds.  Shorter exposures would lead 
to returning to a field too rapidly and hence re-discovering known supernovae, 
while longer exposures would miss supernovae occurring elsewhere.

If it is found that $\alpha \beta$ flattens significantly, tiling will still be 
preferred, but a shorter cadence than the transient duration may be needed to 
avoid observing in increasingly barren fields.  The actual cadence would need to 
be found by noting the new transient accumulation rate in a fresh field, and 
switching to a new field when the rate drops below that in an old field.

The above cadence would {\it only} be valuable for maximizing local supernova 
detections during the rise.  LSST has several other proposed scientific uses, 
however.  Once could length the cadence to optimize for supernovae at higher 
redshifts.  In this light, an LSST Guest Investigator Program might be of 
valuable.  In such a program, scientists outside the LSST collaboration might be 
invited to propose different cadence rates and/or bandpasses so as to optimize 
the detections of sources of different types and/or transients of different 
durations.

\subsection{Maximizing Blazars Discovered with GLAST}

The Gamma Ray Large Area Telescope (GLAST; see, for example, \citealt{Mic02}) 
will surely sample more faint blazers and flares from blazars than ever before 
in the energy ranges from 10 MeV to 100 GeV.  The question has come up, however, 
as to the most productive algorithm for pointing GLAST (J. Bonnell 2002, private 
communication).  The telescope has a planned constraint of pointing away from 
the Earth, so that if the zenith angle of the telescope is not changed, GLAST 
will re-observe the same part of the sky every 90 minutes.  What zenith angle 
rocking algorithm would best maximize the discovery of blazars and blazar 
flares?

From inspection of \citet{Ste96} Figure 2, the power-law slope of 
the cumulative approximated brightness distribution for quiescent blazars is 
expected to be about $-1.3$ below integrated flux in ($>$ 100 MeV photons) of 
$10^{-6}$ cm$^{-2}$ sec$^{-1}$. For flaring blazars, this same power-law slope 
is about $-1.0$.  These estimations are extrapolated from results from the EGRET 
instrument that flew on the Compton Gamma Ray Observatory \citep{Fic96}.

Now GLAST's field of view is about two steradians, and the likely point spread 
function of sources is expected to be highly energy dependent.  The below 
analysis will assume that no matter the energy, sources will not significantly 
overlap, so that source confusion will not significantly flatten the brightness 
distribution.  The energy range for which this will be true may need to be 
determined by actual GLAST observations, but it is assumed valid here through 
most of the GLAST energy band.

First addressed here will be the question of maximizing the number of quiescent 
blazars discovered.  It will be assumed that the background in the gamma-ray 
range of GLAST is dominated by Poisson noise everywhere but in the plane of the 
Galaxy, a relatively small angular region.  The exact boundaries of this region, 
too, will be energy dependent.  Given that $\alpha \beta > 1$ in this region, 
``stare" mode is to be preferred in maximizing the discovery of new blazars.  
This could mean that some zenith angles should be relatively ignored since time 
is better spent re-observing previously observed fields.  Alternatively, a 
``GLAST Deep Field" (GDF) might be created where a significant amount of 
observing time is spent accumulating the relative abundance of dim blazars.  

It is possible, even probable, that $\alpha \beta$ is not constant and will 
flatten significantly for the dimmer quiescents.  In fact, $\alpha \beta$ may 
dip below unity for different exposure times at different galactic latitudes 
and for different energies.  When this happens, tiling becomes preferred, and 
GDFs become inefficient in discovering new blazars.  The most efficient tiling 
algorithm might need to await determination of the actual brightness 
distribution for the fields of interest by GLAST itself.

Next addressed here will be the question of maximizing the number of transient 
blazar flares discovered.  For flares, since $\alpha = -1$, only in the lowest 
noise regimes can $\alpha \beta \ge 1$.  In regions of the sky where the noise 
is entirely dominated by Poisson, the case is a formal tie so that it does not 
matter where in these regions GLAST points.  For everywhere else, however, 
$\alpha \beta < 1$ and so some sort of tiling algorithm will maximize the number 
of flares discovered.  

For regions where tiling is to be preferred, we now address the question of the 
optimal cadence.  Here the situation is complicated by several factors.  The
first is that different regions of the (direction, energy, exposure duration) 
matrix will be best characterized by a different $\alpha \beta$.  The above 
analyses in Section 4 assumed a constant $\alpha \beta$, so that it can only be 
rigorously applied to similar regions.  As indicated above, to determine if it 
is beneficial to jump to a region of different $\alpha \beta$, one should 
determine whether the rate of new transients discovered in the old field has 
dropped below that rate in a fresh field.

For regions of similar $\alpha \beta$, given that down and slew times are 
negligible, the above analyses indicate that the optimal cadence is the duration 
of interest in the blazar flare.  The total duration of blazar flares can be 
from hours to weeks.  The duration of interest may be shorter than this, 
however, if blazar flares will need to be discovered relatively early on, so 
that instruments in other bands can be triggered to monitor the event during the 
flare. 

Here the cadence can be used as tool to isolate blazar flares of a 
given duration -- faster cadences will isolate faster flares.  To best discover 
the fastest blazar flares, GLAST might be put into a mode in which it changes 
its zenith angle rapidly, effectively sampling the entire sky every few hours. 
A caveat occurs for fields away from the spin poles of the Earth.  There a 
cadence of less than 90 minutes is not possible for fields since this is less 
than the revolution time of GLAST around the Earth.

\section{Discussion}

This paper is not meant to be the final word in the determination of pointing 
algorithms for telescope monitors.  Indeed, pointing algorithms will likely need 
to incorporate more practical considerations that are not formally considered 
here.  First, as mentioned above, given a plethora of potential noise sources 
that include cosmic ray hits and satellite glints, it is clearly not assured 
that any single-frame transient is of astronomical importance.  Verification 
observations can and should be built into observing algorithms to assure that 
triggered transients have a reasonable chance of being of astronomical interest.  
When these check observations occur time-contiguous with the initial 
observation, together they can be considered as part what is necessary for 
transient ``discovery."

Next, the idiosyncrasies of different telescopes, observing sites, CCDs, control 
hardware, control software and observers themselves can also have a large and 
even deterministic effect on the design and implementation of a practical 
observing algorithm.  An example of this could be the inability for a telescope 
to slew faster than a certain rate, the need to dither successive observations 
to minimize pixel inequities, or the occurrence of certain fields at certain 
times in areas above cities that create relatively bright sky-glow.

Next, the idiosyncrasies of different transient types can drive practical 
observing algorithms.  Some transients might only be found only in certain 
sections of the sky or superposed on certain types of backgrounds that create 
specific observing challenges.  An example of this is supernova searches, where 
the transient frequently occurs superposed on its host galaxy.  Although 
potentially difficult, this information could be included in the {\it effective} 
$N_{transient}(l)$ function.

Next, the goal of transient observing might not be to discover the maximum raw 
number of transients but the most transients with a certain attribute.  An 
example of this might be an effort to find particularly bright cases of 
microlensing \citep{Nem98} or planet transits \citep{Pep02}.  In these cases, 
observing deeper would not help discover brighter sources.  Here the above 
formalism might be augmented with a weighting function emphasizing sources in 
the desired magnitude range.  Again, alternatively, the effective  
$N_{transient}(l)$ could be adapted to incorporate this information, for example 
not counting transients too dim to be of interest.   

Next, it might be preferred that transients {\it not} be detected in a single 
exposure or time-contiguous series of exposures, as, for example, discussed in 
\citet{Gou99}.  This would greatly affect the discussion given above.  In fact, 
the drive toward separating observing times by the transient duration is fueled 
by the single exposure premise.  If frames can be routinely aligned and co-added 
then one can spread the detection observations out over $t_{dur}$ with any 
distribution at all, only demanding that enough observations be carried out 
during $t_{dur}$ so that transient detection is assured.

Next, transients do not all have the same durations.  Optimizing for a single 
duration might indeed cause an observing algorithm to miss transients of shorter 
duration.  Robust observing algorithms attacking a distribution of durations 
might try to optimize toward the peak of the duration distribution, or use the 
above analysis as the basis for a more sophisticated one that optimizes 
transient discovery rates over the entire distribution of durations.

Nevertheless, even given all of these caveats, transient detection algorithms 
need to be determined more by hard logic and mathematical optimization than by 
whim.  A map of how effective $\alpha \beta$ changes with the accumulated 
exposure time in each field might indeed be useful in matching cadence with 
scientific return.  At minimum, key pieces of information that should be 
considered include the duration and the brightness distributions of the 
transients.  An example of how they interact in a relative clean set of 
theoretical but didactic cases is given above.

Even a valiant effort to predetermine a cadence that optimizes discovery rates 
might fail, given inaccurate knowledge of relevant parameters such as the 
duration distribution function.  Therefore, a pointing algorithm might deploy 
``cadence creep" (M. Kowalski 2003, private communication).  The idea is to 
slightly change cadence over time to see if transient detection rates increase.  
To be effective, enough transients would have to be detected for a statistically 
meaningful comparison.  A search phase for an optimal cadence, between estimated 
boundaries, might be mandated as an early phase of a transient search program.

A natural extension of the idea that different observing algorithms optimize 
different scientific return is the implementation of a diverse array of
observing algorithms on any given sky monitoring telescope. Guest investigator 
programs might diversify previously dedicated sky monitoring telescopes by 
implementing bandpasses and cadences chosen to optimize the discovery of 
different types of transients.

Last, the decision to ``tile" or ``stare" and how fast to tile are influenced by 
more than the ability to discover the maximum amounts of sources and/or 
transients.  The schema discussed above implicitly assumed that other telescopes 
can be deployed for follow-up observations, and that these follow-up telescopes 
will maximize the science uncovered per transient.  If follow-up time is not 
expected for discovered transients, one may want to code follow-up observations 
directly into the timing of the observations.  For this reason, a non-uniform 
cadence, one that combines attributes of both discovery and tracking, such as 
returning to sources in logarithmically increasing time intervals, might be 
preferable. 

I thank the National Science Foundation for support.  I also thank Jerry 
Bonnell, Christ Ftaclas, Andrew Gould, Michael Kowalski, Bohdan Paczynski, J. 
Bruce Rafert, and Thomas Vestrand for helpful comments.

\clearpage
\figcaption[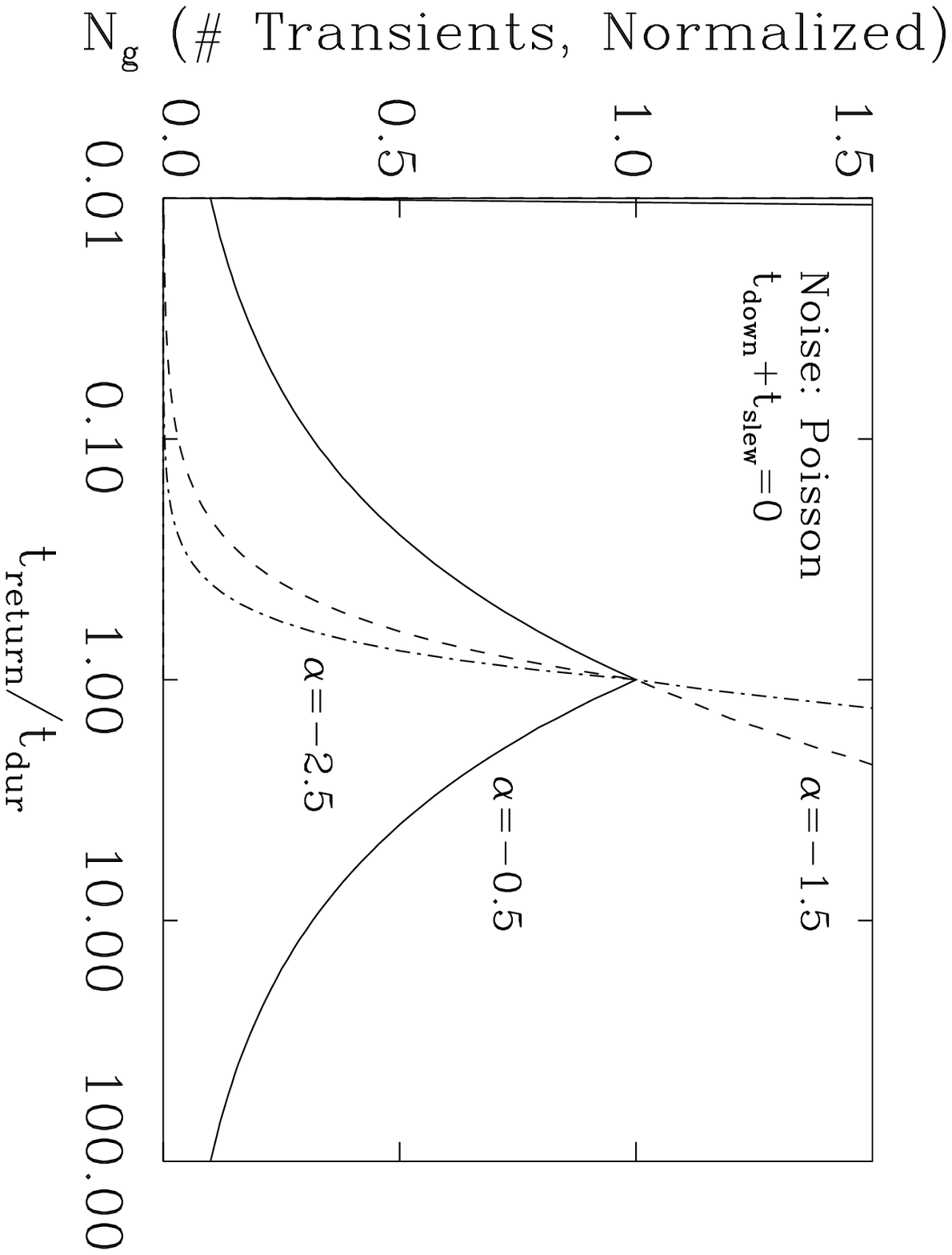]{
A plot of number of transients discovered, $N_g$, versus the time taken to 
return to inspect the same field, $t_{return}$.  $N_g$ has been normalized to 
the number discovered when $t_{return} = t_{dur}$, the duration of the 
transient, while $t_{return}$ is given as a fraction of $t_{dur}$.  Here the 
dominant source of noise is Poisson (``low background", $\beta = -1$), while 
$t_d + t_s$, the down and slew times for the telescope, are taken to be 
negligible. Three power laws of the effective cumulative luminosity distribution 
are depicted: $\alpha=-0.5$ as the solid line, $\alpha=-1.5$ as the dashed line 
and $\alpha=-2.5$ as the dot-dashed line.  For the first two power laws, $N_g$ 
increases monotonically with $t_{return}$, indicating that longer exposures 
detect more transients so that a telescope that ``stares" would discover more 
transients than a similar telescope that ``tiles" the sky.  For the last power-
law, ``tiling" is optimal, while a cadence of $t_{return} = t_{dur}$ maximizes 
the number of transients discovered.
\label{fig1}}

\figcaption[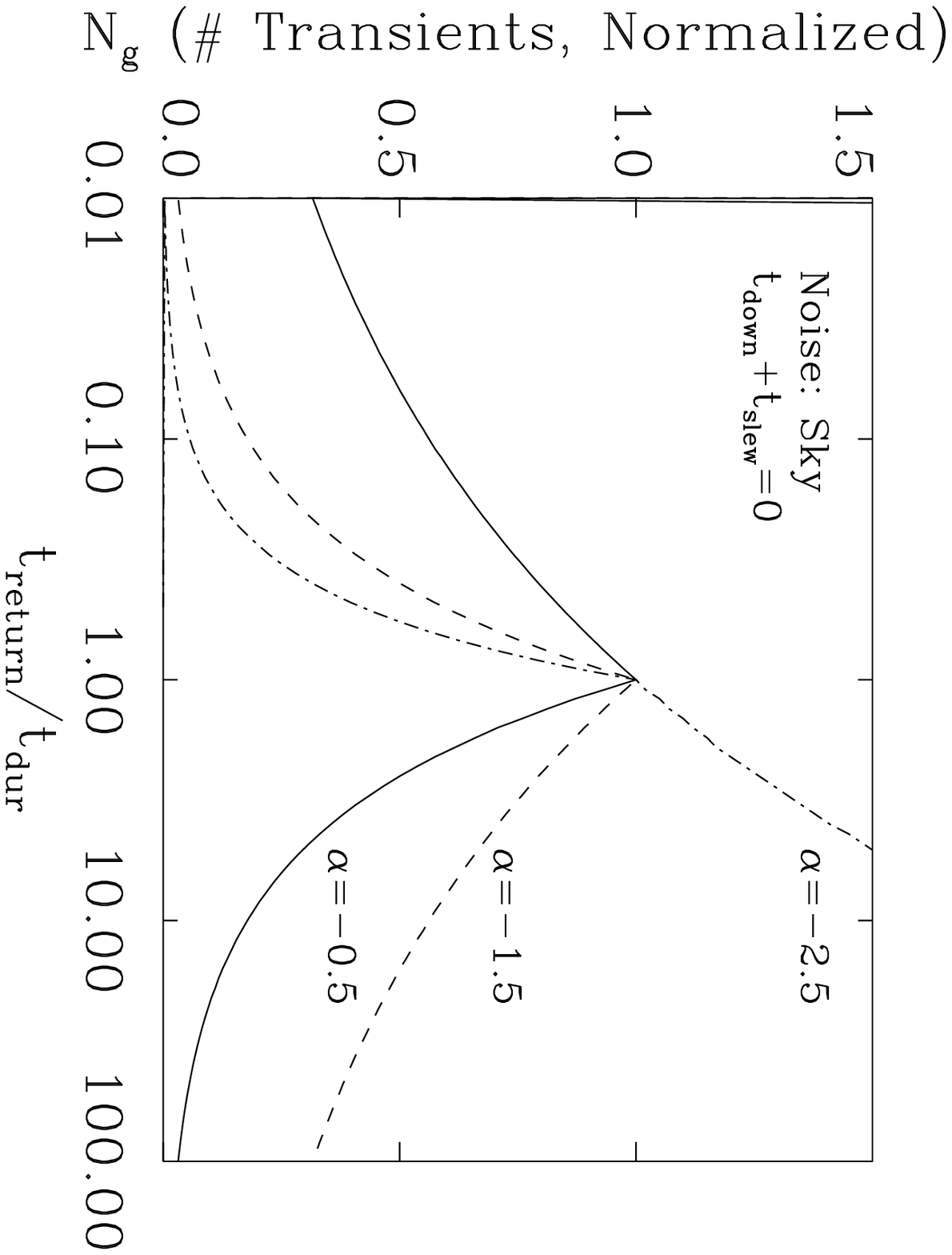]{
Similar to Figure 1 with the exception that the dominant source of noise is 
considered to be sky-glow, the ``high-background" case ($\beta = -0.5$).  Here, 
for the middle $\alpha = -1.5$ case, the most transients are recovered in 
``tile" mode, with the most productive cadence equal to the duration of the 
transient. 
\label{fig2}}

\figcaption[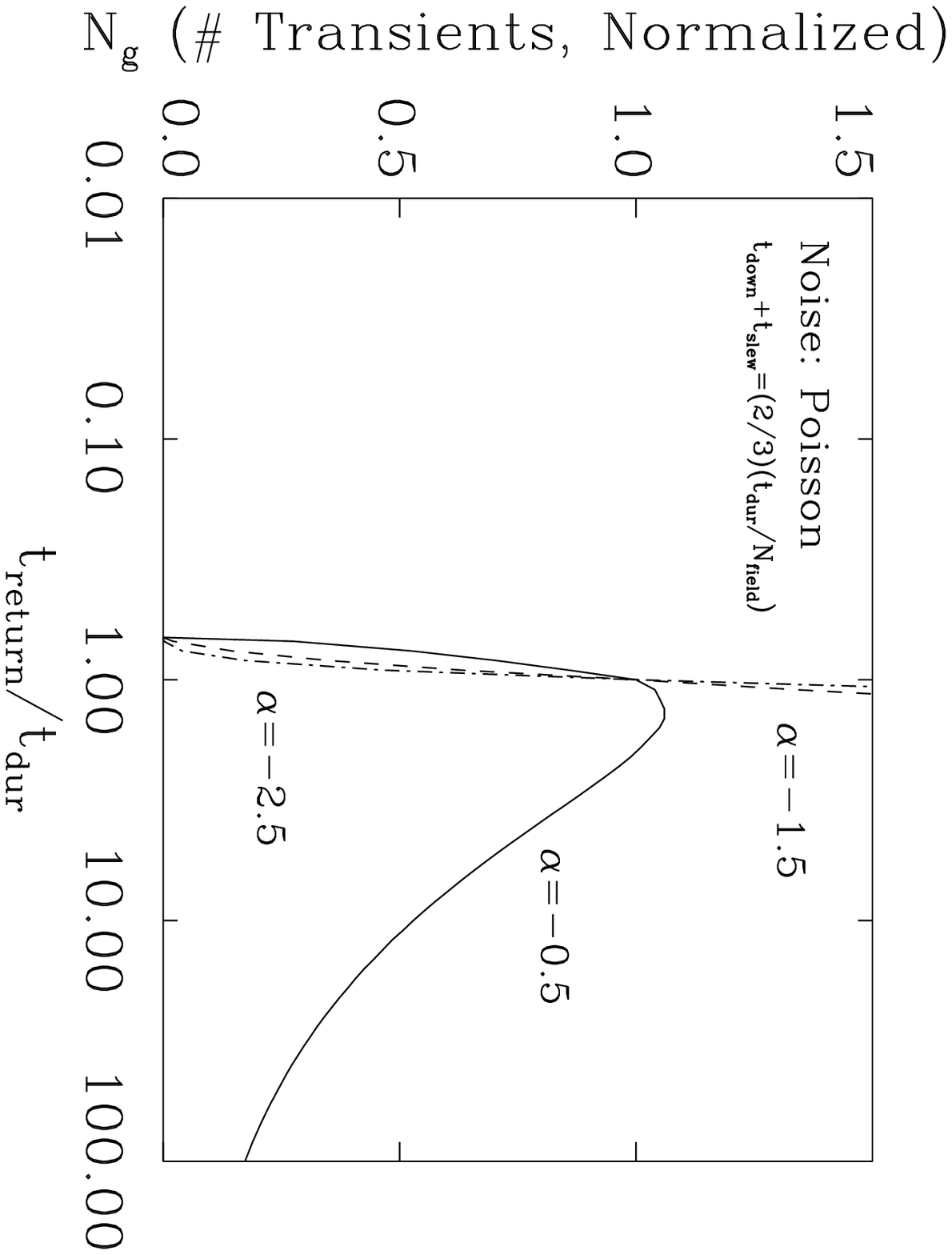]{
Similar to Figure 1 with the exception that significant down plus slew times are 
incurred.  Specifically, $t_d + t_s = (2/3) (t_{dur}/N_{field})$.  Here the  
$\alpha = -0.5$ case recovers the most transients in ``tile" mode, but the 
most productive cadence $t_{return}$ is greater than $t_{dur}$ and determined by 
equation (\ref{treturn}).
\label{fig3}}

\figcaption[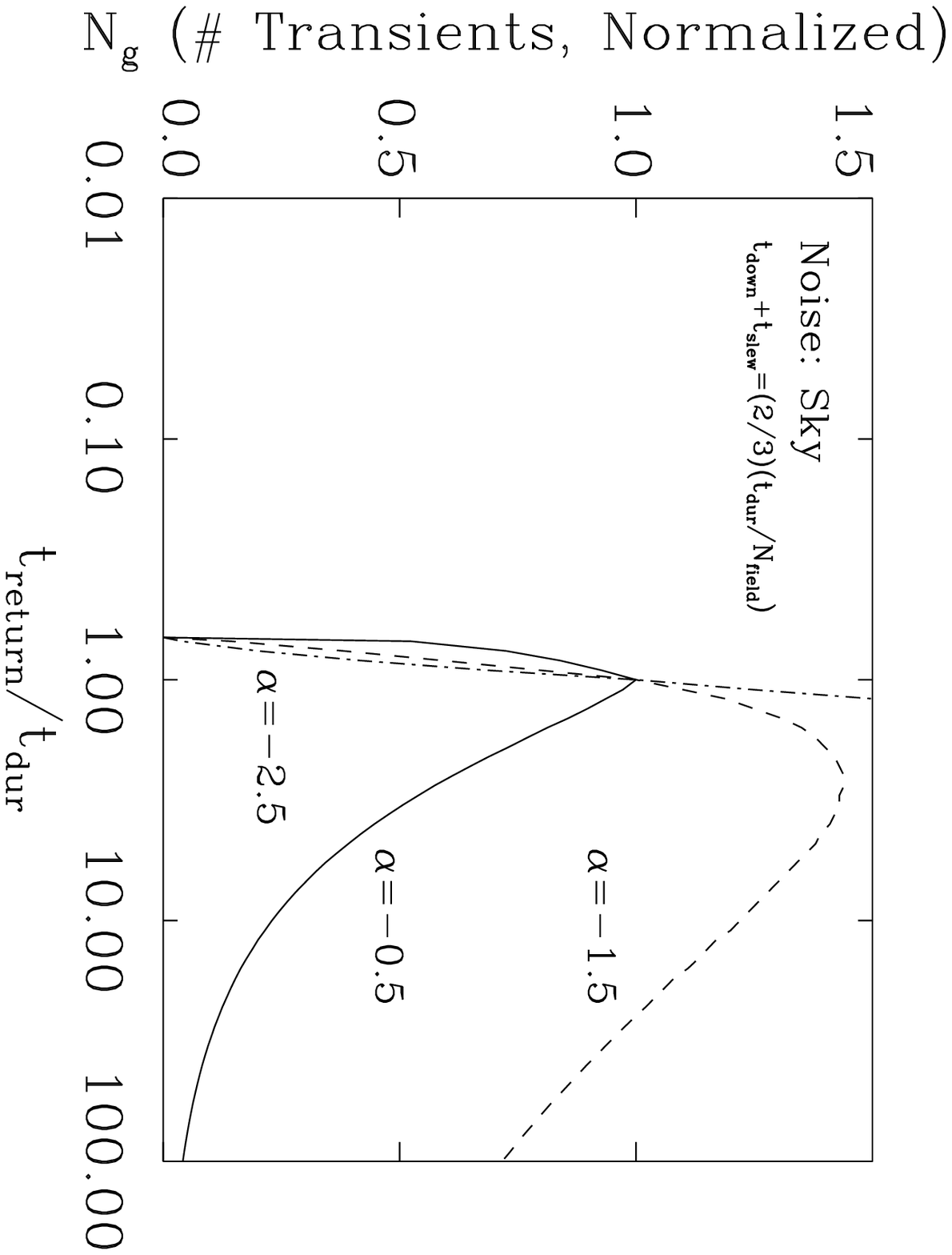]{
Similar to Figure 1 with the exceptions that the dominant source of noise is 
considered to be sky-glow, the ``high-background" case, and significant down 
plus slew times are incurred.  Specifically, $t_d + t_s = (2/3) 
(t_{dur}/N_{field})$.  Here, for the middle $\alpha = -1.5$ case, the most 
transients are recovered in ``tile" mode, with the most productive cadence 
$t_{return}$ being greater than $t_{dur}$ and determined by equation 
(\ref{treturn}).
\label{fig4}}
\newpage

\plotone{Nemiroff.fig1.ps}
\newpage
\plotone{Nemiroff.fig2.ps}
\newpage
\plotone{Nemiroff.fig3.ps}
\newpage
\plotone{Nemiroff.fig4.ps}
\newpage

\end{document}